\documentstyle[12pt,epsfig]{article}
\newlength{\dinwidth}
\newlength{\dinmargin}
\newcommand{\ba}{\begin{array}}
\newcommand{\ea}{\end{array}}
\newcommand{\beq}{\begin{equation}}
\newcommand{\eeq}{\end{equation}}
\newcommand{\bea}{\begin{eqnarray}}
\newcommand{\eea}{\end{eqnarray}}

\def\S{{\bf S}}

\def\bce{\begin{center}}
\def\ece{\end{center}}

\def\nonu{\nonumber}

\def\pa{\partial}
\def\al{\alpha}
\def\be{\beta}

\def\De{\Delta}

\def\La{\Lambda}

\def\S{{\bf S}}

\begin{document}
\thispagestyle{empty}
\addtocounter{page}{-1}
\begin{flushright}
NSF-ITP 99-098\\
SNUST 99-004\\
{\tt hep-th/9908110}\\
\end{flushright}
\vspace*{1.3cm}
\centerline{\Large \bf Three-Dimensional CFTs and RG Flow}
\vskip0.3cm
\centerline{\Large \bf from}
\vskip0.3cm
\centerline{\Large \bf  Squashing M2-Brane Horizon~\footnote{ Work supported 
in part by the U.S. National Science Foundation Grant No. PHY94-07194, KOSEF 
Interdisciplinary Research Grant 98-07-02-07-01-5, KRF International 
Collaboration Grant, and The Korea Foundation for Advanced Studies Faculty 
Fellowship. }}
\vspace*{1.5cm} 
\centerline{\bf Changhyun Ahn${}^a$ {\rm and} Soo-Jong Rey${}^{b,c}$}
\vspace*{1.0cm}
\centerline{\it ${}^a$ Department of Physics, 
Kyungpook National University, Taegu 702-701 Korea}
\vskip0.3cm
\centerline{\it ${}^b$ Physics Department, Seoul National University,
Seoul 151-742 Korea}
\vskip0.3cm
\centerline{\it ${}^c$ Institute for Theoretical Physics, University of 
California, Santa Barbara CA 93106 USA}
\vspace*{0.8cm}
\centerline{\tt ahn@kyungpook.ac.kr, \hskip0.5cm sjrey@phya.snu.ac.kr}
\vskip2cm
\centerline{\bf abstract}
\vspace*{0.5cm}
Utilizing AdS/CFT correspondence in M-theory, an example of interacting $d=3$ 
conformal field theories and renormalization group flow between them is 
presented. Near-horizon geometry of $N$ coincident M2-branes located on a 
conical singularity on eight-dimensional hyperk\"ahler manifold or manifold 
with Spin(7) holonomy is, 
in large-$N$ limit, $AdS_4 \times X_7$, where $X_7$ is seven-sphere
with squashing. Deformation from round $\S_7$ to squashed one is known to
lead to spontaneous breaking of ${\cal N}=8$ local supersymmetry in gauged 
$AdS_4$ supergravity to ${\cal N}=1, 0$. Via AdS/CFT correspondence, 
it is interpreted as renormalization group flow from 
$SO(5) \times SO(3)$ symmetric UV fixed point to $SO(8)$ symmetric 
IR fixed point. Evidences for the interpretation are found both from 
supergravity scalar potential and existence of interpolating static 
domain-wall thereof, and from conformal dimensions of relevant chiral
primary operator at each fixed point.

\vspace*{1.1cm}

\baselineskip=18pt
\newpage

\section{Introduction}
In spacetime of dimensions more than two, few examples are known for
{\sl interacting} conformal field theories or for renormalization group 
flows among themselves. This is especially so in three dimensions, as the
infrared limit corresponds typically to a strong coupling limit. Known examples
include non-Gaussian fixed point of $O(N)$ vector model in large $N$ limit,
infrared limit of (non)abelian gauge theories with ${\cal N} = 
8, 4$ or $ 2$ supersymmetries. The latter can be obtained by 
dimensional reduction of four-dimensional gauge theories with ${\cal N}=4, 
2, 1$ supersymmetries. As such, R-symmetries, holomorphy and supersymmetry
non-renormalization theorems put strong constraints to the
moduli space of these theories. Combining these analysis, a relation called
`mirror symmetry' between 
infrared limit of different gauge theories has been discovered
\cite{mirrorsymmetry} and rederived using brane configurations
\cite{braneconfig} : K\"ahler 
(complex) structure on the 
Coulomb branch of an ${\cal N}=4$ (${\cal N}=2$) gauge theory is mapped to
that on the Higgs branch of its mirror and vice versa. 

In contrast, three-dimensional ${\cal N} = 1$ superconformal field theories
do not have any continuous R-symmetries, holomorphy constraints or any known
non-renormalization theorems. As such, it would be quite difficult to draw any
useful information regarding these theories. The difficulty may be exemplified,
for instance, from the infrared limit of a ${\cal N}=1$ 
brane configuration consisting of $N_c$ D3-branes between a pair of 
rotated NS5-branes and $N_f$ D5-branes: the Higgs branch can be reached from
any point on the Coulomb branch and hence, classically, every point of the
moduli space defines an interacting conformal field theory. The situation 
suggests
that one would really need to follow technically a rather different route 
in order to understand three-dimensional ${\cal N}=1$ superconformal field 
theories.

In this paper, utilizing dual correspondence between
$(d+1)$-dimensional anti-de Sitter supergravity and $d$-dimensional conformal 
field theory (AdS/CFT correspondence) \cite{maldacena, witten, 
gubserklebanovpolyakov}, we will present an example of {\sl interacting} 
${\cal N}=1$ superconformal field theory (with some global symmetries) and 
renormalization group flow thereof. 
The example is based on the well-known result in Kaluza-Klein 
supergravity \cite{s7compactification}:
compactification of eleven-dimensional supergravity to four-dimensional 
anti-de Sitter spacetime on round-$S^7$ with $SO(8)$ isometry or squashed-$S^7$
with $SO(5) \times SO(3)$ isometry \cite{squasheds7}. For the Freund-Rubin 
compactification \cite{freundrubin} 
of eleven-dimensional supergravity \cite{11dsugra}
on a seven-dimensional 
compact manifold $X_7$ that is topologically $S_7$, it has been known that 
round and squashed seven-spheres are the only possibilities, and that both 
are stable, at least, at perturbative level. 
More appropriately, the compactifications correspond to M-theory 
vacua describing near-horizon geometry of two distinct configurations (one 
spherically symmetric and another non-spherical) of $N$ coincident M2-branes. 

In doing so, we will also discover another 
closely related, nonsupersymmetric {\sl interacting} conformal 
field theory, which flows along the same renormalization group trajectory as
the ${\cal N}=1$ counterpart. This comes about as follows. 
An important aspect of the $S_7$ compactification is that, depending on the
choice of the orientation, one and the same squashing deformation leads to 
two distinct vacua . 
Denoting chirality of supercharges on $AdS_4$ (which has descended from 
Killing spinors on $X_7$) as $(N_L, N_R)$, the round and 
the squashed seven-spheres are known to preserve $(8, 8)$ and $(1, 0)$ 
supersymmetries, respectively. 
Because of triality of $SO(8)$, squashing 
deformations can proceed in two different embeddings (so-called 
`skew-whipping' \cite{s7compactification}) of residual isometries:
$SO(8) \rightarrow [SO(5) \times SO(3)]_{\rm c}$ for left-handed orientation 
or 
$[SO(5) \times SO(3)]_{\rm s}$ for right-handed orientation.
Thus, for the squashing with left-handed orientation, the renormalization
group flow interpolates between conformal field theories with ${\cal N}=8$ and 
${\cal N}=1$ supersymmetry, while for the squashing with right-handed 
orientation, the flow interpolates between conformal field theories with 
${\cal N}=8$ and ${\cal N}=0$. 
Along the deformation, the so-called `space invader scenario' -- complicated
level-crossing phenomena among massless and massive Kaluza-Klein states --
then implies that the supersymmetry is broken completely (except the two
endpoints) and the scaling dimension of operators will become renormalized
 in a highly nontrivial manner.  
For both choices of orientation, however, the squashing deformation turns out 
to be governed by one and the same Kaluza-Klein mode. Hence, for both 
${\cal N}=1$ and $0$ conformal field theories, the
renormalization group flow ought to be governed by the same scaling operator. 

We will begin our analysis in section 2 by recapitulating 
relevant aspects of round and squashed $S_7$ compactification vacua in 
eleven-dimensional supergravity, but rephrased in terms of M2-brane parameters.
In section 3, we will investigate `squashing' deformation of each vacua, and
find that the deformation is described by an irrelevant operator at the 
${\cal N}=8$ conformal fixed point, but by a relevant operator at the
${\cal N}=1$ or $0$ conformal fixed points. The renormalization group
flows along the squashing deformation trajectory would then interpolate between
 ${\cal N}=8$ fixed point at the infrared and ${\cal N}=1$ or $0$ at the
ultraviolet. As suggested in \cite{domainwall}, the renormalization group 
flow is described in $AdS_4$ supergravity by a static 
`domain wall' interpolating between the round- and the squashed-$S_7$ vacua. 
In section 4, developing first a general argument regarding nonperturbative
stability, we will reproduce the renormalization group flow along 
the squashing trajectory as a static domain wall. 

Throughout this paper, we will be using the metric convention 
$(-,+,\cdots,+)$. The eleven-dimensional Planck scale is denoted by $\ell_{\rm
p}$.
Our notation is that the $d=11$ coordinates with indices $A, B, \cdots$ are 
decomposed into 
$d=4$ spacetime coordinates $x$ with indices $\alpha, \beta, \cdots$ and
$d=7$ internal space coordinates $y$ with indices $a, b, \cdots$. Denoting
the $d=11$ metric as $\overline{g}_{AB}$ and the antisymmetric tensor field 
as $\overline{F}_{ABCD} = 4 \overline{\nabla}_{[A} \overline{C}_{BCD]}$, 
the bosonic field equations are:
\bea
{1 \over \ell_p^2}
{{\overline R}^A}_B &=& {1 \over 3} \overline{F}^{A PQR} {\overline F}_{BPQR}
- {1 \over 36} \delta^A_B {\overline F}^{PQRS} {\overline F}_{PQRS}
\nonumber \\
\overline{\nabla}_A \overline{F}^{ABCD} 
&=& - {1 \over 4!^2} \overline{\epsilon}^{BCD M_1 \cdots M_8}
\overline{F}_{M_1 \cdots M_4} \overline{F}_{M_5 \cdots M_8}.
\label{11deom}
\eea
\section{${\rm AdS}_4$ Supergravity Vacua: Round and Squashed $\S^7$ }
In this section, we will be recapitulating some of the relevant results 
regarding spontaneous
compactification of $d=11$ supergravity on $AdS_4 \times X_7$, where $X_7$
denotes a seven-dimensional compact Einstein manifold. The $X_7$ refers
to the near-horizon geometry of M2-branes. For $X_7$ diffeomorphic
 to round seven-sphere, it is well-known that there exist only two 
possible Einstein manifolds \cite{squasheds7} \footnote{
By placing M2-branes at conical singularities defined by 
\bea
z_1^2 + z_2^2 + z_3^2 + z_4^3 + z_5^{6k-1} = 0, \hskip1cm
k = {\rm integer}
\nonumber
\eea
it is also possible to get exotic non-spherical horizons that are 
topologically equivalent to $S_7$ but not diffeomorphic to 
it \cite{morrisonplesser}. Isometry group of the 
exotic seven-spheres is $SO(2)$ or $SO(3)$ and hence may correspond to 
${\cal N} = 2, 3$ superconformal field theories.}
: round and squashed seven-spheres. We will denote them by $\S_7$ 
and $\widetilde{\S_7}$, respectively. The latter is a homogeneous space
$[SO(5) \times SO(3)] / [SO(3) \times SO(3)]$ with a weak $G_2$ holonomy.
In fact, generically, M2-branes on a (noncompact) 
eight-dimensional hyperk\"ahler manifold or manifold with $Spin(7)$ holonomy, 
the near-horizon geometry
is expected to vary from spherical to squashed seven-sphere as the branes 
are placed away or at a conical singularity of the manifold \cite{qmw, 
morrisonplesser}. 

Embedded into $d=11$, the interpolating metric between round and squashed
seven-spheres may be written as
\bea
\overline{ds}^2 = R^2 \left[ 
e^{- 7u} g_{\alpha \beta} d x^\alpha d x^\beta
 + e^{2u + 3v} \left( {1 \over 4} d\mu^2 + {1 \over 16} \vec{\omega}^2 
\sin^2 \mu \right)
+ e^{2u - 4v} {1 \over 16} \left( \vec{\nu} + \vec{\omega} \cos \mu \right)^2
 \right].
\label{metric}
\eea
Here, $\vec{\nu}, \vec{\omega}$ are diagonal linear combinations of one-forms,
each satisfying $SU(2)$ algebra:
\bea
\vec{\nu} &=& \vec{\sigma} + \vec{\Sigma}, \quad
\vec{\omega} = \vec{\sigma} - \vec{\Sigma}
\nonumber \\
d \sigma_i &=& -{1 \over 2} \epsilon_{ijk} \sigma_j \wedge \sigma_k
\nonumber \\
d \Sigma_i &=& - {1 \over 2} \epsilon_{ijk} \Sigma_j \wedge \Sigma_k.
\nonumber
\eea
The parameter $R$ measures the overall radius of curvature. The scalar fields 
$u(x)$, 
$v(x)$ parametrize `size' and `squashing' deformation of $S^7$ over $d=4$ 
spacetime: Vol($S^7$) = ${1 \over 3} \pi^4 e^{7u} R^7$, and
squashing is parametrized by $\lambda^2 \equiv e^{- 7v}$. 

Spontaneous compactification of M-theory to $AdS_4 \times S^7$ is 
obtained from near-horizon geometry of $N$ coindicent M2-branes. Geometry of
the horizon can be deformed continuously but,
as mentioned above, the resulting horizon is an Einstein manifold only for 
round- or squashed-$S_7$'s. Through the seven-sphere,  the M2-branes thread 
nonvanishing flux of four-form field strength of the Freund-Rubin form:
\bea
\overline{F}_{\alpha \beta \gamma \delta}
= Q e^{-7 u } \overline{\epsilon}_{\alpha \beta \gamma \delta}
= Q e^{-21 u} \epsilon_{\alpha \beta \gamma \delta}. 
\label{fourform}
\eea
The parameter $Q$ refers to so-called `Page' charge \cite{page}
 $Q \equiv \pi^{-4} 
\int_{X^7} ({}^*F + C \wedge F)$, and is related to the total number of
M2-branes, $N$, as 
\bea
Q = 96 \pi^2 N \ell_{\rm p}^6.
\eea

The $d=4$ field equations resulting from insertion of the ansatz 
Eqs.(\ref{metric}, \ref{fourform}) into Eq.(\ref{11deom}) can be compactly 
summarized by the following effective Lagrangian \cite{page}:
\bea
{\cal L}= \sqrt{-g} \left( R - \frac{63}{2} ( \pa u )^2 -
21 ( \pa v )^2 -V(u, v) \right),
\label{efflag}
\eea
where 
\bea
V(u, v)= e^{-9u} \left[ - 6 e^{+4v} -48 e^{-3v}+12 e^{-10v} 
+  2 Q^2 e^{-12u} \right].
\label{canonicalpotential}
\eea
The field equations are
\bea
\pa^2 u  & = & \frac{6}{7} e^{-9u+4v} +\frac{48}{7} e^{-9u-3v} -
\frac{12}{7} e^{-9u-10v}-
\frac{2}{3} Q^2 e^{-21u}, \nonu \\
\pa^2 v  & = & -\frac{4}{7} e^{-9u+4v} +\frac{24}{7} 
e^{-9u-3v} -\frac{20}{7} e^{-9u-10v},
\label{eom1}
\eea
and
\bea
R_{\alpha \beta} = {63 \over 2} \partial_\alpha u \partial_\beta u
+ 21 \partial_\alpha v \partial_\beta v + g_{\alpha \beta} e^{-9u}
\left[ - 3 e^{+4v} - 24 e^{-3v} +6 e^{- 10v} + Q^2 e^{-12u} \right].
\label{eom2}
\eea
 
The AdS-invariant ground-states correspond to $u, v$ taking constant values
and the spacetime curvature maximally symmetric, $R_{\alpha \beta \gamma
\delta} = {1 \over 3} \Lambda (g_{\alpha \gamma} g_{\beta \delta}
- g_{\alpha \delta} g_{\beta \gamma})$, $R_{\alpha \beta} = \Lambda g_{\alpha
\beta}$. The aforementioned two vacua of Eqs.(\ref{eom1},\ref{eom2}) are:
\bea
\S_7 \,\, \quad : \quad u = u_1 &=& {1 \over 12} \ln (3^{-2} Q^2 ),
\qquad v = v_1 = 0 \qquad \qquad(\lambda^2 = 1)
\nonu \\
\Lambda_1 &=& -12 \left\vert {Q \over 3} \right\vert^{-3/2}
\label{roundparameter}
\eea
and
\bea
\widetilde{\S_7} \quad : \quad 
u = u_2 &=& {1 \over 12} \ln (3^{-4} 5^{10/7} Q^2 ), \qquad
v = v_2 = {1 \over 7} \ln 5 \qquad \left(\lambda^2 = {1 \over 5} \right), 
\nonu \\
\Lambda_2 &=& - 12 \cdot 3^{7/2} 5^{-5/2}
\left\vert {Q \over 3} \right\vert^{-3/2},
\label{squashedparameter}
\eea
respectively. 

\begin{figure}[htb]
\label{fig1}
\vspace{0.5cm}
\epsfysize=8cm
\epsfxsize=8cm
\centerline{
\epsffile{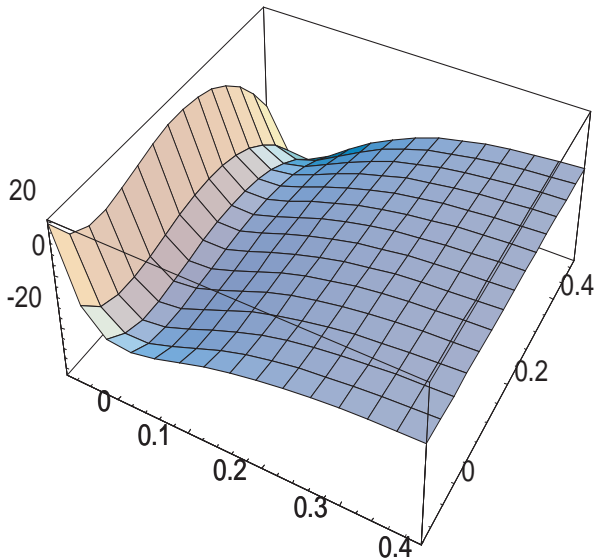} }
\vspace{0.5cm}
\caption{\sl Scalar potential $V(u, v)$ for $Q = 3$. The local 
minimum point and the saddle point correspond to the $\S_7$ and the 
$\widetilde{\S_7}$, respectively. Note that the steepest descent
is nearly along the $v$ (squashing parameter) direction. Both ground-states
are stable against $u$ (breathing) direction deformation.}
\end{figure}
\vspace{0.5cm}

The scalar potential $V(u, v)$ depicted in Figure 1 displays the two
critical points: $\S_7$ is a saddle point and corresponds to a minimum 
(nearly) along $v$-direction, while the $\widetilde{\S_7}$ is 
a maximum \footnote{
Numerically, $\sqrt{-\Lambda_2} = 0.91..\sqrt{-\Lambda_1}$.}. 
Recall that $u, v$ fields parametrize all possible deformation of $S^7$, but 
restricted to retain at least $SO(5) \times SO(3)$ isometry subgroup. 
The $\S_7$ critical point is actually invariant under the full $SO(8)$
isometry group, while the $\widetilde{\S_7}$ is invariant only under 
the minimal $SO(5) \times SO(3)$ subgroup.

The $\S_7$ has a trivial holonomy group that it gives rise to 
$(N_L, N_R) = (8, 8)$ Killing spinors. Therefore, for either choice of the
orientation of seven-sphere, the near-horizon geometry preserves maximal 
${\cal N}=8$ supersymmetry. 
On the other hand, having a weak $G_2$ holonomy, it turns out that the 
$\widetilde{\S_7}$ 
gives rise to $(1, 0)$ Killing spinors. As such, near-horizon
geometry preserves minimal ${\cal N}=1$ supersymmetry for left-handed 
orientation of the seven-sphere and no supersymmetry at all for right-handed
orientation. The two alternative choices of the seven-sphere orientation
may be viewed as reversal of Page charge, which is identical to the M2-brane
charge (up to numerical factors). This means that, when approaching a  
conical singularity of an eight-dimensional manifold with Spin(7) holonomy, 
M2- and $\overline{\rm M2}$-branes behave differently. In particular, there
will generically 
be a net attractive force among $\overline{\rm M2}$-branes \footnote{
In fact, conical singularities of hyperk\"ahler or Calabi-Yau fourfolds 
exhibit the same phenomena. The near-horizon geometry $X_7$ is, for each
of them, 3-Sasaki manifold with Killing spinors $(3, 0)$ or 
Sasaki-Einstein manifold with Killing spinors $(2, 0)$, 
respectively \cite{qmw}.}.

\section{Three-Dimensional Conformal Field Theories}
In the investigation of the Kaluza-Klein compactification on the
seven-sphere, the pheonomenon of `squashing' has been interpreted as 
triggering spontaneous (super)symmetry breaking via the (super) Higgs-Kibble
mechanism \cite{s7compactification}. 
Two important features discovered regarding the compactification and 
`squashing' deformation thereof are the `space invader scenario' and 
`skew-whipping' orientation reversal. 
According to the `space invader scenario', along the continuous
`squashing' deformation, Kaluza-Klein spectrum for spin 2, 1, and $0^+$
do not change under the `skew-whipping'. In particular, the `squashing'
deformation is part of $0^+$ scalar modes and hence it can be studied
on equal footing for both orientation of the seven-sphere.
 
In this section, utilizing the above aspects and the results of section 2 
on Kaluza-Klein spectrum under squashing deformation, we will be 
identifying an operator
that gives rise to a renormalization group flow associated with the symmetry 
breaking $SO(8) \rightarrow SO(5) \times SO(3)$ and find that the operator is 
relevant at the $\widetilde{\S_7}$ fixed point but becomes irrelevant at the
${\S_7}$ fixed point. 

\subsection{$SO(8)$ Invariant Conformal Fixed Point}
Let us begin with $X_7 = \S_7$. Associated with it is the $d=3$ conformal 
field theory with ${\cal N}=8$ supersymmetry and $SO(8)$ R-symmetry. This
theory describes the conformal origin of the moduli space of three-dimensional
${\cal N}=8$ supersymmetric Yang-Mills theory with gauge group $SU(N)$, 
at which the gauge coupling becomes infinitely strong. Away from the origin, 
the gauge group is broken to $U(1)^{N-1}$ and the moduli space is given by
${\bf R}^{8N}/ S_N$ \cite{16susy}. 

To identify conformal field theory operator corresponding to the `squashing' 
perturbation (while preserving $SO(5) \times SO(3)$ isometry), we shall
be considering harmonic fluctuations of spacetime metric and $u(x), v(x)$ 
scalar fields around $AdS_4 \times \S_7$. Inferring from Eq.(5), fluctuation 
of these fields is compactly summarized by the Lagrangian:
\bea
{\cal L}_{\S_7}= \sqrt{-g} \left( R - 2\La_1-\frac{63}{2} ( \pa u )^2 -
21 ( \pa v )^2 -2V_1(u, v) \right),
\label{roundfluctuation}
\eea
where $\Lambda_1 = {1 \over 2} 
V(u_1, v_1)$ is the $AdS_4$ cosmological constant and 
the scalar potential $V_1(u, v)$ is given by
\bea
V_1(u, v)=-\La_1 \left( 1-\frac{1}{4} e^{-9(u-u_1)} ( e^{4v}+8e^{-3v}-
2e^{-10v} ) +\frac{3}{4} e^{-21(u-u_1)} \right).
\eea

For our present purpose, following \cite{page}, it turns out more convenient 
to rewrite Eq.(\ref{roundfluctuation}) in terms of the un-rescaled, 
M-theory metric $\overline{g_{\al \be}} = e^{-7u} g_{\al \be}$:
\bea
{\cal L}_{\S_7} & = & \sqrt{-\overline{g}} e^{7u} \left( \, \overline{R} - 
105 ( \pa u )^2 -
21 ( \pa v )^2 -e^{7u} V(u, v) \, \right) \nonu \\
& =  & \sqrt{-\overline{g}} e^{7u} \left( \, \overline{R} -
2\overline{\La_1} - 
105 ( \pa u )^2 - 21 ( \pa v )^2 -2 \overline{V_1(u, v)} \, \right).   
\eea 
Here, the scalar potential is 
\bea
\overline{V_1(u, v)}= 
-\overline{\La_1} \left( 1-\frac{1}{4} e^{-2(u-u_1)} ( e^{4v}+8e^{-3v}-
2e^{-10v} ) +\frac{3}{4} e^{-14(u-u_1)} \right),
\eea
in which the un-rescaled cosmological constant $\overline{\La_1}
= e^{7u_1} \La_1$ is given by 
\bea
\overline{\La_1} \equiv - 12 m_1^2 {1 \over \ell_p^2}= 
-12 \left( \frac{|Q|}{3} \right)^{-1/3} {1 \over \ell_p^2}
\qquad {\rm where} \qquad m_1 = {1 \over \overline{r_{\rm IR}}   }.
\eea
We have denoted the curvature radius as $\overline{r_{\rm IR}}$ since, 
as we shall
see shortly, the round ${\S_7}$ ground-state is stable against `squashing'
perturbation and hence corresponds to an infrared stable fixed point.
Moreover, comparing the cosmological constant with that of spherical, 
near-horizon geometry of $N$ coincident M2-branes, one finds that 
$\overline{r_{\rm IR}}$ is related to $N$ and Planck scale $\ell_{\rm p}$ as 
$\overline{r_{\rm IR}} = \ell_{\rm p} {1 \over 2} ( 32 \pi^2 N)^{1/6}$.

Conformal dimension of the perturbation operator that represents the
`squashing' is determined by fluctuation spectrum of the scalar fields. 
After rescaling the scalar fields as $\sqrt{210} u \equiv \overline{u}, 
\sqrt{42} v \equiv \overline{v}$, one finds that the (correctly normalized)
fluctuation spectrum for $v$-field around the $\S_7$ takes a positive value:
\bea
M_{vv}^2 (\S_7) 
= \left[ \frac{\pa^2} {\pa \overline{v}^2} 2 \overline{V_1}
\right]_{\overline{u}=\overline{u_1}, \overline{v}=\overline{v_1}=0} = 
\qquad -
\frac{4}{3} \overline{\La_1} \ell_p^2= + 16 m_1^2.
\label{roundvmass}
\eea

The $v$-field represents `squashing' of $\S_7$ and hence, under $SO(8)$
isometry group, ought to correspond to ${\bf 300}$, the lowest mode of the 
transverse, traceless symmetric tensor representation. Recall that, on
$\S_7$, mass spectrum of the representation corresponding to $SO(8)$
Dynkin label $( {\bf n-2, 2, 0, 0})$ is given by
\bea
\widetilde {M^2}   = \left( (n + 3)^2 - 9 \right) m^2,
\eea
where $m^2$ is mass-squared parameter of a given $AdS_4$ spacetime
and mass of a scalar field $S$ is defined according to 
$(\De_{\rm AdS} + \widetilde{ M^2} ) S = 0$. This follows easily from the 
known mass formula \cite{bcers} $M^2 = \De_L - 4 m^2$, where the 
Lichnerowicz operator $\De_L$ has eigenvalues $\De_L = [n(n+6) + 12]m^2$ for
$O^{+(2)}$, and the fact that $M^2$ is traditionally defined 
according to $(\De_{\rm AdS} - 8 m^2 + M^2 ) S = 0$. For ${\bf 300}$ 
(corresponding to ${\bf n} = 2$), $\widetilde{ M^2}_{\bf 300} = 16 m^2$ and
this ought to equal to Eq.(\ref{roundvmass}). 

Indeed, recalling that $\overline{r_{\rm IR}} =\frac{1}{2} l_{\rm p} 
( 32 \pi^2 N )^{1/6}$, one finds 
that Eq.(\ref{roundvmass}) fits perfectly with ${\bf 300}$ spectrum:
\bea 
M_{vv}^2 (\S_7) = 16 \left( {\vert Q \vert \over 3} \right)^{-1/3}
\quad = \quad {16 \over \overline{r^2_{\rm IR}}} = {\widetilde{M_{\bf 300}}}^2.
\eea
Via AdS/CFT correspondence, one thus concludes that, in $d=3$ conformal field 
theory with ${\cal N} = 8$ supersymmetry, the $SO(5) \times SO(3)$ symmetric 
`squashing' ought to be an irrelevant perturbation of conformal dimension 
$\Delta = 4$. Note that this is the same for either choice of the 
seven-sphere orientation.

An important point is that the `squashing' deformation arises not at the
lowest level but at the second of the $S_7$
Kaluza-Klein tower. As such, the deformation probes M-theory beyond the 
so-called consistent truncation of eleven-dimensional supergravity, viz.
four-dimensional ${\cal N} = 8$ gauged supergravity. 

\subsection{$SO(5) \times SO(3)$ Invariant Conformal Fixed Point}
Consider next the conformal fixed point corresponding to the
$X_7 = \widetilde{ \S_7}$. As mentioned already, due to the possibility of 
`skew-whipping', the fixed point could be either left-squashed 
$\widetilde{\S_{\rm 7L}}$ with ${\cal N}=1$ supersymmetry or right-squashed 
$\widetilde {\S_{\rm 7R}}$ with no supersymmetry. 

From Eq.(5), harmonic fluctuation of $u(x), v(x)$ scalar fields around $X_7
= \widetilde{ \S_7}$ is compactly summarized by the Lagrangian:
\bea
{\cal L}_{\widetilde \S_7} = \sqrt{-g} 
\left( R - 2\La_2-\frac{63}{2} ( \pa u )^2 -
21 ( \pa v )^2 -2V_2(u, v) \right),
\label{squashedlagrangian}
\eea
where the scalar potential $V_2(u, v)$ is given by
\bea
V_2(u, v) =  -\La_2 \left( 1-\frac{1}{36} e^{-9(u-u_2)} 
\left( 25 e^{4(v-v_2)}+40 e^{-3(v-v_2)}-
2 e^{-10(v-v_2)} \right)  +\frac{3}{4} e^{-21(u-u_2)} \right).
\eea
Again, in terms of the un-rescaled M-theory metric, the Lagrangian 
Eq.(\ref{squashedlagrangian}) may be reorganized as
\bea
{\cal L}_{\widetilde \S_7}  = 
\sqrt{-\overline{g}} e^{7u} \left( \, \overline{R} -
2\overline{\La_2}  - 
105 ( \pa u )^2 -
21 ( \pa v )^2 -2 \overline{V_2(u, v)} \, \right),   
\eea 
where 
\bea
\overline{V_2(u, v)}  = 
 -\overline{\La_2} \left( 1-\frac{1}{36} e^{-2(u-u_2)} 
\left( 25 e^{4(v-v_2)}+ 40 e^{-3(v-v_2)}- 2 e^{-10(v-v_2)} \right)  
+\frac{3}{4} e^{-14(u-u_2)} \right),
\eea
and the un-rescaled cosmological constant $\overline\La_2 =
e^{7 u_2} \La_2$ is given by
\bea
\overline{\La_2}  &\equiv& - 12 m_2^2 {1 \over \ell_{\rm p}^2}
= - 12 \cdot 3^{7/3} 5^{-5/3}  \left( {\vert Q \vert  \over 3} \right)^{-1/3} 
{1 \over \ell_{\rm p}^2}, \quad {\rm where}
\quad m_2 = {1 \over \overline{ r_{\rm UV}}   }.
\eea
Once again, the scalar field $v(x)$ parametrize `squashing' perturbation
around $\widetilde{\S_7}$. Mass spectrum of the $v(x)$ field is calculated 
straightforwardly:
\bea
M^2_{vv}[\widetilde{ \S_7}] \equiv \left[ \frac{\pa^2}
{\pa \overline{v}^2} 2 \overline{V_2} \right]_{\overline{u}=
\overline{u_2}, \overline{v}=\overline{v_2}} \,\, = \,\, + 
\frac{20}{27} \overline{\La_2} \ell^2_{\rm p} = -\frac{80}{9} m_2^2.
\label{mass2}
\eea
The spectrum is tachyonic but stays above the Breitenlohner-Freedman bound
\cite{breitlohnerfreedman} 
and hence, in the corresponding dual, unitary conformal field theory, 
the `squashing' deformation ought to be described by a relevant operator. 

The tachyonic spectrum Eq.(\ref{mass2}) can be understood as follows.
Under $SO(8) \rightarrow SO(5) \times SO(3)$ , the 
branching rule of a $SO(8)$ Dynkin label $({\bf 0, 2, 0, 0})$ (corresponding
to the representation ${\bf 300}$) in terms of $SO(5)$ Dynkin label 
$({\bf p, q})$ and $SO(3)$ Dynkin label ({\bf r}) (or in terms of 
their respective representations) is given as follows \cite{branching}:
\bea
 {\bf (0, 2, 0, 0 )} 
 & = & {\bf (02)(4) \oplus (21)(2) \oplus (40)(0) \oplus (01)(4)
\oplus (02)(0)} \nonu \\
&& {\bf \oplus (20)(2)  \oplus (01)(2) \oplus (00)(4) \oplus (00)(0)}. 
\eea
In terms of their representations,
\bea
{\bf 300} & = & {\bf (14,5) \oplus (35,3) \oplus (35,1) \oplus (5, 5) 
\oplus (14, 1)} 
\nonu \\
& & {\bf \oplus (10,3) \oplus (5,3) \oplus (1, 5) \oplus (1, 1).   }
\label{branching}
\eea
Since the `squashing' preserves $SO(5) \times SO(3)$ isometry group, 
the spectrum Eq.(\ref{mass2}) ought to correspond to that of the singlet
(the last state in the branching rule, Eq.(\ref{branching})). One can check
this explicitly. 
The eigenfunctions of the Lichnerowicz operator $\De_L$ appear in the 
$SO(5) \times SO(3)$ representations found by decomposing the ${\bf 
(0, 2, 0, 0)}$ of $SO(8)$. Depending on the particular representations of
$SO(5) \times SO(3)$, eigenvalues are given by \cite{duff}:
\bea
\De_L= \frac{20}{9} m^2  \left(C_G +\frac{9}{5} \right) \qquad \mbox{or}
\qquad \frac{20}{9} m^2  \left( C_G + \frac{8}{5} \pm \frac{2}{\sqrt{5}}
\sqrt{C_G+\frac{1}{20}} \right)
\label{dl}
\eea
for ${\bf 8}_{\rm v}$, ${\bf 8}_{\rm s}$ and ${\bf 8}_{\rm c}$, 
respectively. Here,
$C_G$ denotes the second-order Casimir operator for the isometry group 
and is given by $C_G= C_{SO(5)} +3 C_{SO(3)}$. Recall that, in our notation,
mass-squared spectrum of a $O^{+(2)}$ scalar field $S$ satisfying 
$(\De_{\rm AdS} + \widetilde{M^2} ) S = 0$ is given by $\widetilde{M^2} = 
\De_L - 12 m^2$. For the singlet $({\bf 1, 1})$, $C_G=0$ and one obtains
${\widetilde{M_{\bf (1,1)}}}^2 = -8 m^2 \mbox{ or}  -\frac{80}{9} m^2 $. 
Henceforth, identifying Eq.(\ref{mass2}) with the lower, one identifies
\bea
M_{vv}^2 (\widetilde{\S_7})
= - {80 \over 27} \left(3^{-11} 5^5 \vert Q \vert \right)^{-1/3}
\qquad = \qquad
- {80 \over 9} {1 \over \overline{r^2_{\rm UV}} } 
= \widetilde{M_{\bf (1, 1)}}^2.
\label{massspect}
\eea
One thus finds that the perturbation that corresponds to `squashing'
around $X_7 = \widetilde{ \S_7}$ has a scaling dimension either 
$\Delta = 4/3$ or $5/3$ and hence corresponds to a relevant operator.

Note that, of the two possible scaling dimensions, the lower one is 
below the naive unitarity bound $3/2$ in three-dimensional conformal
field theories. The two-fold ambiguity of the scaling dimension arises 
for relevant operators whose scaling dimension is below the naive 
unitarity bound and, in $AdS_4$ spacetime, originates from the
fact that a scalar field whose mass is in the range
$(-9/4, -5/4)$ (in units of $4m^2$) can be quantized in two inequivalent
boundary conditions \cite{breitlohnerfreedman, balakraus}. 
Klebanov and Witten \cite{klebanovwitten} 
have argued that the two possible choices of
the scaling dimension are related to two distinct dual conformal field 
theories and have confirmed it explicitly in several examples in which
R-symmetry is continuous. In the present situation, by the same argument,
the two scaling dimensions of `squashing' deformation operator ought to 
correspond to each of the two ${\cal N} = 1, 0$ conformal field theories,
but, as there is no continuous R-symmetry, we were not able to resolve
the twofold ambiguity completely.

\section{Squashing Domain Wall and RG Flows} 
In the previous section, round and squashed seven-spheres have been identified
as two possible near-horizon geometry of $N$ coincident M2-branes. Both 
seven-spheres then 
represent stable vacua of eleven-dimensional supergravity, at 
least in perturbative expansion in powers of $1/N$ and $\ell_{\rm p}$, as 
there is no tachyon mode equal to or below the Breitenlohner-Freedman 
bound \cite{breitlohnerfreedman}
\footnote{Nonsupersymmetric vacua obtained out of skew-whipping are always
stable. This has been proven in \cite{berkoozrey}.}. 
Are the two vacua stable
also nonperturbatively? We will find that, surprisingly enough, an answer to 
this question turns out to be intimately connected to the holography of 
anti-de Sitter spacetime, existence of {\sl static} domain-wall, and 
renormalization group flow between fixed points. 

We have seen that, in Eqs.(\ref{roundparameter}, \ref{squashedparameter}), 
since $\La_1 < \La_2$, the $\widetilde{\S_7}_{\rm L, R}$ 
ground-states have higher potential 
energy than the $\S_7$ ground-state. The aforementioned perturbative
stability states that, starting from the $\widetilde{\S_7}$ ground-state, 
there 
cannot be any smooth roll-over along the steepest descent to the $\S_7$ one.
On the other hand, nonperturbatively, one might suspect that there may be a 
potential source of instability: the $\widetilde{\S_7}$ ground-state, which 
is apparently a false vacuum of the gauged supergravity potential, may 
`tunnel' (without barrier), as a result of false vacuum decay \cite{colemandeluccia}, to the 
$\S_7$ ground-state. 

\begin{figure}[htb]
\label{fig2}
\vspace{0.5cm}
\epsfxsize=12cm
\epsfysize=7.5cm
\centerline{
\epsffile{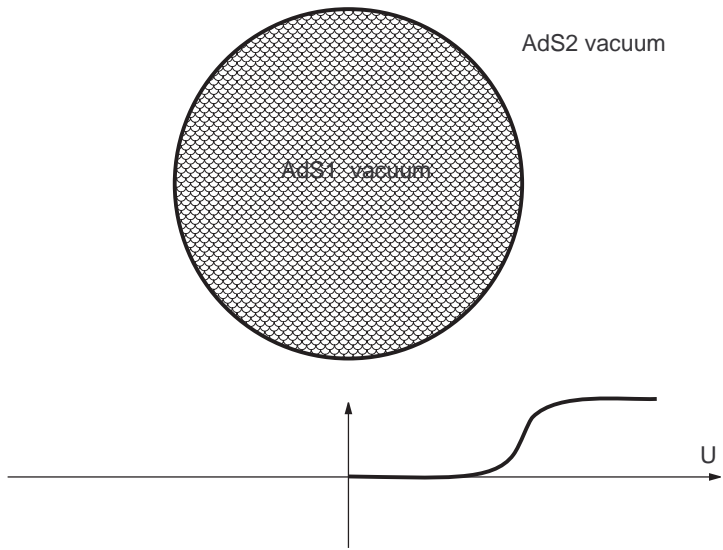}
           }
\vspace{0.5cm}
\caption{\sl 
Schematic view of static domain wall interpolating the two vacua of 
$AdS_4$ supergravity. Note that interpolating scalar fields 
are critically damped between the two vacua.}
\end{figure}
\vspace{0.5cm}

Semiclassically, such a tunnelling process would be described by a
gravitational instanton. However, utilizing the result of 
\cite{weinberg, cveticgriffiesrey}, 
it is straightforward to see that the relevant gravitational instanton 
has an infinite action, leading to a complete suppression of the false vacuum
decay. Consider an 
ansatz of the gravitational instanton in $AdS_4$ that preserves,
at the least, the $SO(3)$ rotational symmetry. It then follows that unique 
instanton configuration is a time-translation invariant bubble configuration -- a bubble of the $\S_7$ vacuum surrounded
by the  $\widetilde{\S_7}$ vacuum. As the instanton is translationally 
invariant along time direction, the semiclassical
instanton amplitude is zero. Analytically continuing to Minkowski spacetime
(i.e. $AdS_4$), one now obtains a {\sl static} $SO(3)$ symmetric domain-wall
interpolating between the $\S_7$ and $\widetilde{\S_7}$ vacua. We will now
analyze the configuration.

Consider the equations of motion Eq.(4) for $g_{\al \be}, u, v$. Let us 
take the following boost-invariant (viz. manifestly $Poin(2,1)$ invariant
along the domain-wall) ansatz:
\bea
u &=& {1 \over \sqrt{63}} \, {\tilde u}(z), 
\quad v = {1 \over \sqrt{42}} \, {\tilde v}(z), 
\nonumber \\
ds^2_4 &=& {r^2 \over z^2} \left( dz^2 + e^{2h(z)} \eta_{\alpha \beta} 
dx^\alpha d x^\beta \right), \qquad \eta_{\alpha \beta} = (-++).
\label{ansatz}
\eea
The ansatz is subject to boundary conditions:
\bea
ds^2_4 &\rightarrow&  \frac{r^2_{\rm IR}}{z^2} \left( dz^2 + 
\eta_{\alpha \beta} dx^\alpha dx^\beta \right), \;\;\; \, 
u \rightarrow u_1, \quad \, v \rightarrow v_1 \, \qquad (z \rightarrow \infty)
\nonumber \\
d s^2_4 &\rightarrow& {r^2_{\rm UV} \over z^2} \left( dz^2 + 
\eta_{\alpha \beta} dx^\alpha dx^\beta \right),
\quad u \rightarrow u_2, \quad v \rightarrow v_2 \qquad (z \rightarrow 0)
\label{boun}
\eea
after appropriate change of variables is made. Note that the boundary 
conditions are manifestly boost invariant.

Components of Ricci tensor and Ricci scalar in the background 
Eq.(\ref{ansatz}) are given by
\bea
R_{00} & = & + \, \frac{3}{t^2} \, \left( -1+ th' - t^2h'^2-t^2h'' \right), 
\nonu \\
R_{11} & = & -\frac{e^{2h}}{t^2} \left( -3 + 5th' -3t^2h'^2 -t^2h''\right), 
           \quad R_{22} = R_{33} = - R_{11} 
\nonu \\
R & = & -\frac{6}{ r^2} \left( 2 -3th'+2 t^2h'^2+t^2h'' \right).
\eea
Denoting $\partial_z = {}'$, the field equations of motion are given by
\bea
{\tilde u}'' +\left( 3h' -\frac{2}{z} \right) {\tilde u}' - 
\frac{r^2}{z^2} \frac{\pa V}{\pa {\tilde u}} =0,
\label{ueom}
\\
{\tilde v}''+\left( 3h' -\frac{2}{z} \right) {\tilde v}' - 
\frac{r^2}{z^2} \frac{\pa V}{\pa {\tilde v}} =0,
\label{veom}
\\
2 h''+ 2 \frac{h'}{z} + {1 \over 2} \left( 
{\tilde u}'^2 + {\tilde v}'^2 \right) =0,
\label{heom}
\\
6 \left ( h'- {1 \over z} \right)^2 
    - {1 \over 2} \left( {\tilde u}'^2 + {\tilde v}'^2 \right) 
+ \frac{r^2}{z^2} V =0.
\label{inteom}
\eea

One immediately notes that Eqs.(\ref{ueom}, \ref{veom}) may be interpreted
as equations of motion of an `analog' particle (of unit mass) in two 
dimensions, whose coordinates are  parametrized by $\tilde{u}$ and $\tilde{v}$,
under the influence of an `analog' potential $-V(\tilde{u}, \tilde{v})$.
The particle is also subject to a frictional force. The `dynamic friction 
coefficient' $(3 h' - 2/z)$ turns out negative-definite always. Intuitively, this follows
from the fact that $h' = 0$ in the limit the gravity is turned off 
$\ell_{\rm p } \rightarrow 0 $ and that the physics ought to be analytic in
powers of $\ell^2_{\rm p}$. Because of the anti-friction, the `analog' particle 
can start from the minima of $-V$ initially (viz. at $z = 0$) and then creep 
up to the maxima of $-V$ finally (viz. at $z = \infty$).
What is less clear is that the anti-frictional energy injected to the `analog' 
particle equals precisely to the potential energy difference $2 (\Lambda_1 
- \Lambda_2 )$.

Answer to the question is provided by the fact that, from Eqs.(32-33), 
$\tilde{u}, \tilde{v} \rightarrow 0$ is satisfied at critical points
of the scalar potential $V$ and the fact that the resulting Minkowski 
domain-wall is an extremal configuration, saturating
domain-wall energy {\sl density}. An important point is that this also applies
to the nonsupersymmetric $\widetilde \S_{\rm 7R}$ vacuum, 
as the field equations 
Eqs.(\ref{ueom} - \ref{inteom}) govern stability for both 
$\widetilde\S_{\rm 7L}$ and $\widetilde\S_{\rm 7R}$. 
The extremality condition implies that energy gain by creating a 
region of true vacuum around $z = 0$ inside false vacuum is balanced exactly 
by mass of the $SO(3)$ symmetric domain-wall, viz. saturation of the 
Coleman-DeLuccia bound and hence is {\sl a posteriori} 
consistent with the static domain-wall ansatz. 
While an exact solution of the domain-wall configuration is not possible in
analytic form, asymptotics of the domain-wall can be studied straightforwardly.
Using the asymptotics, we will now check consistency of the $SO(3)$ symmetric
domain-wall configuration. In the IR region, $z \rightarrow \infty$, 
asymptotic solutions of $h(z)$, $\tilde{u}(z)$ and $\tilde{v}(z)$ are given by
\bea
h(z) & \sim & 0 +  \frac{h_\infty}{z^{2a}} +\cdots,  \nonu \\
\tilde{u}(z) & \sim & 
\sqrt{7 \over 16} \ln ( 3^{-2} Q^2) + {\tilde{u}_\infty \over z^a} + \cdots
\nonu \\
\tilde{v}(z) & \sim &  0 + \frac{\tilde{v}_\infty}{z^a} +\cdots
\nonumber
\eea 
Solving the equations of motion Eqs.(\ref{ueom} - \ref{inteom}), one obtains
\bea
16 h_\infty + \left( \tilde{u}_\infty^2 + \tilde{v}_\infty^2 \right) 
&=& 0, \nonumber \\
3 + r^2 \Lambda_1 &=& 0.
\nonumber
\eea
The first equation implies that
$h_\infty < 0$, viz. the scale factor $h(z)$ asymptotes to $0^-$. Then,
using this fact and the rescaling relation $\overline{r^2_{\rm IR}} 
= e^{7 u_1} {r^2_{\rm IR}}$, 
one finds that the second equation implies that $r = r_{\rm IR}$. 

In the UV region, $z \rightarrow 0$, asymptotic solutions of $h(z)$,
$\tilde{u}(z)$, and $\tilde{v}(z)$ are given by
\bea
h(z) &\sim& (1 - \lambda) \ln z + {h_0 z^{2b}} + \cdots
\nonu \\
\tilde{u}(z) &\sim& \sqrt{7 \over 16} \ln \left( 3^{-4} 5^{10/7} Q^2 \right)
+ {\tilde{u}_0  z^b} + \cdots
\nonu \\
\tilde{v}(z) &\sim& 
\sqrt{42 \over 7} \ln 5 + {\tilde{v}_0  z^b} + \cdots.
\nonumber
\eea
One again finds that 
\bea
\lambda^2 = 
\frac{V(u_2, v_2)}{V(u_1, v_1)}
=\frac{\La_2}{\La_1}=
\frac{3^{7/2}}{5^{5/2}}  < 1.
\label{asym}
\eea  
Change of variables $w = z^c$ and trivial rescaling of $x^\alpha$ coordinates
bring the asymptotic metric back into the $AdS_4$ form, but with a scaled
radius of curvature $r_{\rm IR} / \lambda $. This is precisely the 
radius of curvature $r_{\rm UV}$ of the $\widetilde{\S_7}_{\rm L,R}$ vacua,
and hence confirms the consistency of the asymptotic solution. From the
equations of motion, one also finds conditions $h_0 < 0$ and 
 $b^2 + ({3 \over 2} \lambda
- 2) b - {5 \over 26} \lambda^2 = 0$. Choosing positive root of $b$, one
obtains regular asymptotote of Eq.(\ref{asym}) with monotonic rescaling
of the curvature of radius.

Another question is during the interpolation between the two vacua, whether 
the scalar fields $u, v$ are critically damped (as depicted in Figure 2) or
under-damped (in which case the scalar fields will excute finite number of
oscillations before settling down to the maxima of $-V$). In the `analog 
particle' interpretation, mode decomposition (obtained from Laplace transform 
of Eqs.(33, 34) around minimum of $-V$) takes precisely the same form as 
the Fourier transform of their equations of motion in $AdS_4$. 
Suppose that the scalar field is tachyonic below the 
Breitenlohner-Freedman bound. In this case, two eigen-frequencies take a 
complex-value and corresponds to an under-damped interpolation.  
Hence, the interpolation is always critically damped for scalar fields whose
masses are above the Breitenlohner-Freedman bound.  
Restated, renormalization group flows between unitary 
conformal field theories ought to be monotonic always and never exhibits 
an oscillatory behavior.

\begin{figure}[htb]
\label{fig3}
\vspace{0.5cm}
\epsfysize=8cm
\epsfxsize=10cm
\centerline{
\epsffile{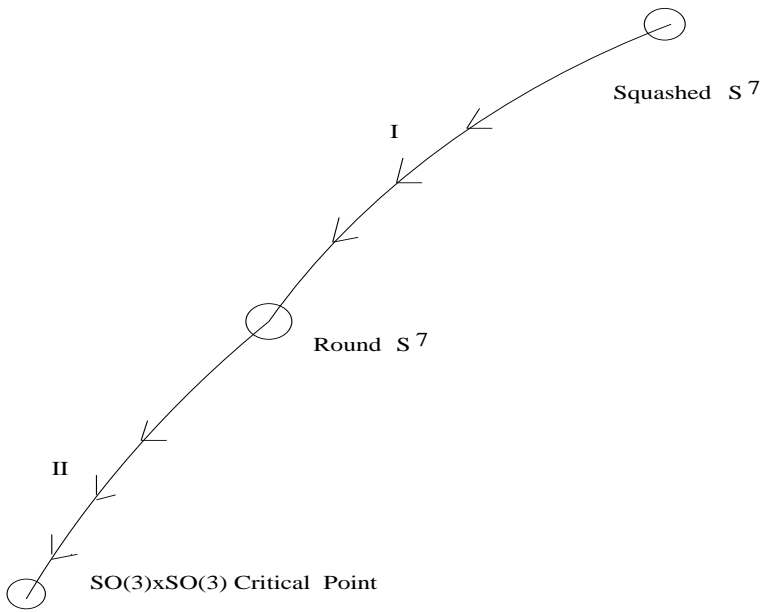}
           }
\vspace{0.5cm}
\caption{\sl The RG flow along the $SO(5)\times SO(3)$ invariant direction 
for I: $\widetilde{\S^7} \rightarrow $ $\S^7$ and along 
the $SO(3) \times SO(3)$
invariant direction for II: $SO(8) \rightarrow SO(3) \times SO(3)$.
}
\end{figure}
\vspace{0.5cm}

The monotonic radial behavior of the static $SO(3)$ domain-wall configuration
is, as suggested by \cite{domainwall}, the holographic representation
of the renormalization group flow and a version of $c$-theorem thereof.
We have shown that well-defined renormalization-group flow
and $c$-theorem exist only if the supergravity vacua are stable both
perturbatively and nonperturbatively. Nonperturbative instability of a 
perturbatively stable vacuum is signalled by an existence of a gravitational
instanton with a {\sl finite} Euclidean action. Analytically continuing 
to Minkowski spacetime, the instanton corresponds to an expanding (or 
contracting) domain-wall instead of being a static one. Under the AdS/CFT 
correspondence, such a time-dependent process does not 
admit interpretation as a renormalization group flow. 

So far, we have considered a particular one-parameter renormalization group
flow between
conformal fixed points of M2-brane worldvolume theory. Geometrically,
the flow is induced from varying the position of M2-brane when placed 
near a conical singularity of an eight-dimensional manifold with 
$Spin(7)$ holonomy. In the infrared limit, the conical singularity and
hence squshing of M2-brane horizon are washed out completely. At the
infrared fixed point with $SO(8)$ symmetry, one may also flow further into 
another fixed points by turning on a set of relevant operators. It includes 
scalar operators of Dynkin label $({\bf n}, 0, 0, 0), \,\, {\bf n} \ge 2$
and pseudoscalar operators of Dynkin label $({\bf n}, 0, 2, 0), \,\, 
{\bf n} \ge 0$. Among them are 70 scalar fields ${\bf 35}_{\rm v} \oplus
{\bf 35}_{\rm c}$ of $SO(8)$ in the massless gravity supermultiplet,
parametrizing the coset space $E_{7(7)}/SU(8)$. 
Decomposing them under $SO(5) \times SO(3) \subset SO(8)$,
${\bf 35}_{\rm v} + {\bf 35}_{\rm c} \rightarrow
2 [ ({\bf 5, 1}) + ({\bf 10, 3}) ]$. Turning on the two relevant operators
$({\bf 5, 1})$ breaks $SO(8) \rightarrow SO(3) \times SO(3)$. Utilizing
on the known result \cite{warner}, Distler and Zamora \cite{distlerzamora}
have studied renormalization group flow to a nonsupersymmetric vacuum 
with $SO(3) \times SO(3)$ global symmetry (See Figure 3). 
Unlike the `skew-whipping'
nonsupersymmetric vacua as we have studied, however, their analysis does not
shed any light on stability of the vacua. It is a logical possibility that, 
along a direction in coset space $E_{7(7)}/SU(8)$ (
orthogonal to $SO(3) \times SO(3)$ orbit), perturbations
violate the Breitenlohner-Freedman bound and the corresponding conformal
field theory is nonunitary.  

We would like to thank D.R. Morrison, I. Klebanov, M. Porrati and E. Witten 
for useful discussions. We acknowledge warm hospitality of Institute des 
Hautes Scientifique (SJR),  Institute for Theoretical Physics at Santa Barbara
(SJR), and Center for Theoretical Physics at Seoul National University (CA),
where part of this work was undertaken.

\end{document}